\documentclass[apj]{emulateapj}

\usepackage{apjfonts}
\usepackage{natbib}
\usepackage{mathptmx}

\newcommand{\ltsim}{\raisebox{-.5ex}{$\;\stackrel{<}{\sim}\;$}}
\newcommand{\gtsim}{\raisebox{-.5ex}{$\;\stackrel{>}{\sim}\;$}}
\newcommand{\kms}{\ifmmode {\rm km\ s}^{-1} \else km s$^{-1}$\fi}
\newcommand{\vFWHM}{\ifmmode V_{\mbox{\tiny FWHM}} \else
            $V_{\mbox{\tiny FWHM}}$\fi}
\newcommand{\msun}{$M_{\odot}$}
\newcommand{\et}{et al.\ }

\newcommand{\hb}{H$\beta$}
\newcommand{\mbh}{$M_{\rm BH}$}
\newcommand{\nvciv}{\ion{N}{5}/\ion{C}{4}}

\shortauthors{SHEMMER ET AL.}
\shorttitle{A METALLICITY--ACCRETION RATE RELATIONSHIP IN AGNs.}

\journalinfo{The Astrophysical Journal, 614:???--???, 2004 October 10,
astro-ph/0406559}

\slugcomment{Received 2004 March 28; accepted 2004 June 22}

\begin{document}

\title{Near Infrared Spectroscopy of High Redshift Active Galactic Nuclei. \\
I. A Metallicity--Accretion Rate Relationship}
\author{
O.~Shemmer,\altaffilmark{1}
H.~Netzer,\altaffilmark{1}
R.~Maiolino,\altaffilmark{2}
E.~Oliva,\altaffilmark{3}
S.~Croom,\altaffilmark{4}
E.~Corbett,\altaffilmark{4}
and L.~di Fabrizio \altaffilmark{3}
}

\altaffiltext{1}
                {School of Physics and Astronomy and the Wise
                Observatory, The Raymond and Beverly Sackler Faculty of
                Exact Sciences, Tel-Aviv University, Tel-Aviv 69978,
                Israel; ohad@wise.tau.ac.il, netzer@wise.tau.ac.il}

\altaffiltext{2}
                {INAF - Osservatorio Astrofisico di Arcetri, Largo E. Fermi 5,
                I-50125 Firenze, Italy; maiolino@arcetri.astro.it}

\altaffiltext{3}
                {Istituto Nazionale di Astrofisica, Centro Galileo Galilei, and
                Telescopio Nazionale Galileo, P.O. Box 565, E-38700
                Santa Cruz de la Palma, Spain; oliva@tng.iac.es}

\altaffiltext{4}
                {Anglo-Australian Observatory, PO Box 296, Epping, NSW
                1710, Australia; scroom@aaoepp.aao.gov.au}

\begin{abstract}
We present new near infrared spectroscopic measurements of the \hb\ region for
a sample of 29 luminous high redshift quasars. We have measured the
width of \hb\ in those sources, and added archival \hb\ width measurements,
to create a sample of 92 active galactic nuclei (AGNs) for which \hb\
width and rest-frame UV measurements of \ion{N}{5}~$\lambda1240$ and
\ion{C}{4}~$\lambda1549$ emission-lines are available. Our sample
spans six orders of magnitude in luminosity and includes 31
radio-loud AGNs. It also includes 10 narrow-line Seyfert 1 galaxies
and one broad absorption-line quasar.
We find that metallicity, indicated by the \nvciv\ line ratio, is primarily
correlated with accretion rate, which is a function of luminosity and \hb\
line-width. This may imply an intimate relation between starburst, responsible
for the metal enrichment of the nuclear gas, and AGN fueling, represented
by the accretion rate.
The correlation of metallicity with luminosity, or black hole (BH) mass,
is weaker in contrast with recent results which were based on measurements of
the width of \ion{C}{4}. We argue that using \ion{C}{4} as a proxy to \hb\
in estimating \mbh\ might be problematic and lead to spurious BH mass
and accretion rate estimates in individual sources. We discuss the potential
implications of our new result in the framework of the starburst--AGN
connection and theories of BH growth.
\end{abstract}


\section{Introduction
\label{introduction}}

The determination of chemical abundances in active galactic nuclei (AGNs)
provides a powerful tool to probe star forming activity in galactic nuclei
at present, and in the early Universe. Early metal abundance studies in AGNs
utilized weak and broad intercombination lines,
such as \ion{N}{3}]~$\lambda1750$, \ion{N}{4}]~$\lambda1486$, and
\ion{O}{3}]~$\lambda1663$, to determine
metal abundances in the broad-line region (BLR) gas (Shields 1976). These
early studies have found that BLR metallicity is about solar, with some
luminous quasars reaching super-solar values.
Hamann \& Ferland (1993; hereafter HF93) found that emission-line ratios,
such as \ion{N}{5}~$\lambda1240$/\ion{C}{4}~$\lambda1549$ (hereafter \nvciv)
can trace the BLR metallicity across a wide
range of luminosity and BLR density (see also Hamann \et 2002 for
a more detailed model). HF93 also found that BLR metallicity, determined
in this way, correlates with luminosity, in what they proposed as the
metallicity--luminosity ($Z-L$) relationship in AGNs. Given the well known
relationship between AGN luminosity and black hole (BH) mass ($M_{\rm BH}$),
HF93 suggested that the $Z-L$ relationship can naturally lead to a 
$Z-M_{\rm BH}$ dependence, in analogy with the mass--metallicity relation
observed in some elliptical galaxies (e.g., Trager \et 2000).

Shemmer \& Netzer (2002; hereafter SN02) have shown that once
narrow-line Seyfert 1 galaxies (NLS1s) are introduced to the $Z-L$ diagram,
they deviate significantly from the $Z-L$ relation by exhibiting high
\nvciv\ at low luminosity. A possible explanation
is that \nvciv\ is not an adequate metallicity indicator for
NLS1s (and perhaps other AGNs). Otherwise, this implies that at least at
low-luminosity, the $Z-L$ relation is more complex and cannot be a simple
two-parameter dependence.
According to SN02, BLR metallicity also depends on the width of \hb,
which is perhaps the best accretion rate (in terms of the Eddington ratio:
$L_{\rm Bol}/L_{\rm Edd}$) indicator, and that this dependence may prevail
also at the high-luminosity end. To test this hypothesis, we observed the
near infrared (IR) spectrum of a sample of 29 luminous $2<z<3.5$ quasars,
with available \nvciv\ data, in order to measure the width of their \hb\
lines, and thus deduce the accretion rate in these sources. 

This paper is part of a two-paper project describing near-IR spectroscopic
measurements of luminous high-redshift quasars aimed at estimating
$L_{\rm Bol}/L_{\rm Edd}$ and measuring the emission line properties in such
objects.
In this (first) paper we describe the quasars' selection, their observations
and data reduction (\S~\ref{observations}). A description of the data analysis
is given in \S~\ref{analysis}. In \S~\ref{results} we summarize our main
results concerning the broad emission lines, and in \S~\ref{discussion} we
discuss their potential implications. A summary of our main conclusions is
given in \S~\ref{conclusions}. A following paper (Netzer \et 2004; hereafter
Paper II) addresses the complementary issue of the [\ion{O}{3}] and \ion{Fe}{2}
lines in those sources.

\begin{deluxetable*}{lccccccl}
\tablecolumns{8}
\tablewidth{0pt}
\tablecaption{Observations Log
\label{log}}

\tablehead
{
\colhead{Quasar Name} &
\colhead{RA (2000)} &
\colhead{DEC (2000)} &
\colhead{$z$\tablenotemark{a}} &
\colhead{$z_{\rm sys}$\tablenotemark{b}} &
\colhead{$H$\tablenotemark{c}} &
\colhead{Observatory} &
\colhead{Date} \\
\colhead{(1)} &
\colhead{(2)} &
\colhead{(3)} &
\colhead{(4)} &
\colhead{(5)} &
\colhead{(6)} &
\colhead{(7)} &
\colhead{(8)}
}
\startdata
2QZ J001221.1$-$283630     & 00 12 21.1 & $-$28 36 30 &2.327&{\bf 2.339}&{\bf
  16.6}& AAT & 2003 Jul 17 \\
2QZ J002830.4$-$281706     & 00 28 30.4 & $-$28 17 06 &{\bf 2.406}&2.401&{\bf
  15.9}& AAT & 2003 Jul 16 \\
UM 667                     & 00 47 50.1 & $-$03 25 31 &3.122&3.132& \nodata
& TNG & 2003 Aug 18 \\
LBQS 0109$+$0213           & 01 12 16.9 & $+$02 29 48 &2.343&2.349&{\bf 15.3}&
TNG & 2002 Nov 8 \\
\protect{[HB89]} 0123$+$257\tablenotemark{d}& 01 26 42.8 & $+$25 59 01 &2.358
&2.369&{\bf 15.9}& TNG & 2002 Aug 27 \\
2QZ J023805.8$-$274337     & 02 38 05.8 & $-$27 43 37 &{\bf 2.452}&{\bf 2.471}
&{\bf 16.0}& AAT & 2003 Jul 15 \\
SDSS J024933.42$-$083454.4 & 02 49 33.4 & $-$08 34 54 &2.491&2.491& 16.5
& AAT & 2003 Jul 16 \\
\protect{[HB89]} 0329$-$385\tablenotemark{d}& 03 31 06.3 & $-$38 24 05 &2.423
&2.435&{\bf 15.6}& AAT & 2003 Jul 14 \\
\protect{[HB89]} 0504$+$030\tablenotemark{d}& 05 07 36.4 & $+$03 07 52 &2.463
&2.473& \nodata & TNG & 2002 Nov 10 \\
SDSS J100428.43$+$001825.6 & 10 04 28.4 & $+$00 18 26 &3.046&3.046& \nodata
& TNG & 2003 Nov 13 \\
TON 618\tablenotemark{d}   & 12 28 24.9 & $+$31 28 38 &2.219&{\bf 2.226}&
{\bf 13.9}& TNG & 2003 Jun 4 \\
\protect{[HB89]} 1246$-$057& 12 49 13.9 & $-$05 59 19 &2.236&{\bf 2.240}&
{\bf 14.3}& TNG & 2003 Jun 6 \\
\protect{[HB89]} 1318$-$113& 13 21 09.4 & $-$11 39 32 &2.308&2.306&{\bf 15.1}&
TNG & 2003 Jun 5 \\
\protect{[HB89]} 1346$-$036& 13 48 44.1 & $-$03 53 25 &2.344&{\bf 2.370}&
{\bf 15.1}& TNG & 2002 Apr 5 \\
SDSS J135445.66$+$002050.2 & 13 54 45.7 & $+$00 20 50 &2.511&{\bf 2.531}&
\nodata & TNG & 2004 Apr 1 \\
UM 629                     & 14 03 23.4 & $-$00 06 07 &2.462&2.460&{\bf
  16.0}& AAT & 2003 Jul 16 \\
UM 632\tablenotemark{d}    & 14 04 45.9 & $-$01 30 22 &2.518&2.517&{\bf 16.1}&
TNG & 2003 Jun 5 \\
UM 642                     & 14 10 26.4 & $-$00 50 09 &2.372&2.361&{\bf 16.6}&
TNG & 2003 Jun 4 \\
UM 645\tablenotemark{d}    & 14 11 23.5 & $+$00 42 53 &2.269&2.257& \nodata
& TNG & 2003 Jun 7 \\
SBS 1425$+$606             & 14 26 56.1 & $+$60 25 50 &3.160&3.202&{\bf 14.5}&
TNG & 2003 Aug 30 \\
SDSS J170102.18$+$612301.0 & 17 01 02.2 & $+$61 23 01 &2.293&{\bf 2.301}&
{\bf 16.4}& TNG & 2003 Jun 4 \\
SDSS J173352.22$+$540030.5 & 17 33 52.2 & $+$54 00 30 &3.425&3.428&{\bf 15.7}&
TNG & 2003 Aug 30 \\
\protect{[HB89]} 2126$-$158& 21 29 12.2 & $-$15 38 41 &3.268&3.282&{\bf 14.9}&
TNG & 2003 Oct 15 \\
\protect{[HB89]} 2132$+$014\tablenotemark{d}& 21 35 10.6 & $+$01 39 31 &
3.194&3.199& \nodata & TNG & 2003 Sep 5 \\
2QZ J221814.4$-$300306     & 22 18 14.4 & $-$30 03 06 &{\bf 2.384}&2.389&{\bf
  16.0}& AAT & 2003 Jul 16 \\
2QZ J222006.7$-$280324     & 22 20 06.7 & $-$28 03 24 &2.406&2.414&{\bf
  14.3}& AAT & 2003 Jul 15 \\
\protect{[HB89]} 2254$+$024\tablenotemark{d}& 22 57 17.5 & $+$02 43 18 &
2.081&2.083&{\bf 15.9}&
TNG & 2003 Jun 16 \\
2QZ J231456.8$-$280102     & 23 14 56.8 & $-$28 01 02 &{\bf 2.392}&2.400& 16.6
& AAT & 2003 Jul 17 \\
2QZ J234510.3$-$293155     & 23 45 10.3 & $-$29 31 55 &2.360&2.382& 16.5
& AAT & 2003 Jul 17 \\
\enddata
\tablenotetext{a}{Redshift taken from the NASA/IPAC Extragalactic Database
(NED; http://nedwww.ipac.caltech.edu/), except for {\it bold face} values
that were taken from the 2QZ archive.}
\tablenotetext{b}{Systemic redshift determined from the peak of the
[O {\sc iii}]~$\lambda5007$ emission line. In cases where no
[O {\sc iii}] emission was detected, H$\beta$ was used instead
({\it bold face}).}
\tablenotetext{c}{Magnitudes in bold face are 2MASS measurements,
while others were obtained by photometric measurements using field comparison
stars, whose magnitudes were taken from the 2MASS archive. Six sources
do not have 2MASS magnitudes (or $H$ images with comparison stars having 2MASS
magnitudes), and their rest-frame optical fluxes were obtained from
extrapolation of the rest-frame $\lambda1450$ flux.}
\tablenotetext{d}{Radio loud quasar.}
\end{deluxetable*}

\section{Sample Selection, Observations, and Data Reduction}
\label{observations}

We have selected a sample of 29 luminous quasars for our study upon the
following criteria:

\begin{enumerate}
\item Luminous ($L\gtsim 10^{46}$ erg s$^{-1}$) sources with expected $H$
magnitudes $\ltsim17$ to allow high-S/N spectra on medium size
telescopes.
\item Each source has an archived or published rest-frame UV spectrum
that includes the \ion{N}{5} and \ion{C}{4} emission lines, which are
not subject to severe absorption.
\item \hb\ is located in the observable IR bands, where it is
least affected by atmospheric absorption.
\end{enumerate}

According to these criteria all of our quasars lie in a $2<z<3.5$ redshift
range, with \hb\ located in either the $H$ or $K$ bands. Observations were
done at the Anglo-Australian Telescope (AAT) in Australia and at Telescopio
Nazionale Galileo (TNG) in Spain. An observations log and quasar details
appear in Table~\ref{log}.

The observations at TNG were done with the Near Infrared Camera Spectrometer
(NICS). Eight quasars were observed with a $1\arcsec$-wide slit, and
eleven with a $0.5\arcsec$-wide slit.
During the observations the telescope was nodded along the slit and
each spectrum was taken at 2--6 different positions along the spatial
axis in order to perform the primary background subtraction.
We used the HK grism to obtain the
$1.4-2.5~\mu$m wavelength range, which provides an almost uniform
dispersion of 11.2 \AA /pix throughout this range. This results in a
resolving power of ${\rm R}\sim500$ (1\arcsec\ slit) or ${\rm R}\sim1000$
(0.5\arcsec\ slit), slightly increasing towards the red.
Each night, spectra of one or two standard O stars were taken as close as
possible in time and in airmass to the quasars to allow the removal of
telluric absorption features from the quasars' spectra.

The spectroscopic observations at AAT were done with the second generation
Infrared Imager \& Spectrograph (IRIS2) and a $1\arcsec$-wide slit.
During the observations the telescope was nodded along the slit in an
ABBA sequence, thus each spectrum was broken into a set of 4 frames
to allow primary background subtraction.
We used the Sapphire-H Grism together with an $H$ filter to obtain the
$1.54-1.90~\mu$m wavelength range\footnote {Normally, this setup allows
a coverage of the $1.485-1.781~\mu$m range, i.e., the entire $H$
band, however, a misplacement of the grism in the filter wheel during our
run, resulted in a short wavelength cutoff.}, with a dispersion
of 3.4 \AA /pix, and a resolution of ${\rm R}\sim2500$.
Standard stars of spectral types O, B, and G were observed each night
to remove telluric features from the quasars' spectra. 
We also used the imaging mode of IRIS2 to take $H$ band images of our
quasars. $H$ magnitudes for the quasars were obtained by relative
photometry and comparison with $H$ magnitudes of nearby stars taken
from the two micron all sky survey (2MASS) archive
\footnote {http://www.ipac.caltech.edu/2mass/}.
Seven of the quasars also had $H$ magnitudes in the 2MASS archive.
Non-systematic differences of up to $\sim0.5$ mag were found between
these magnitudes and the magnitudes obtained by the relative photometry. This
may be mainly due to increased photometric errors near the edge of
the 2MASS detection limit.

\begin{figure*}
\plotone{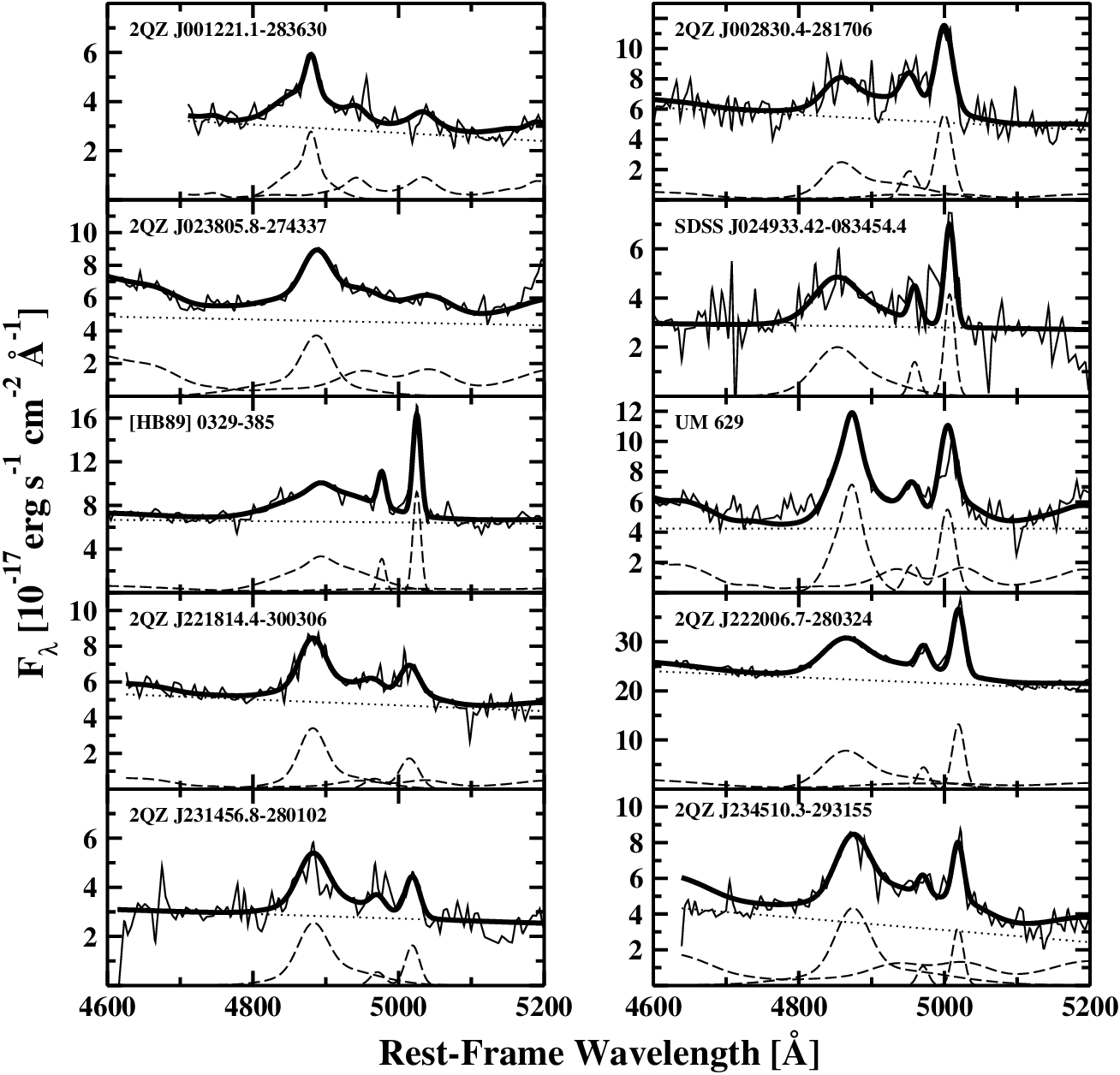}
\caption
{IR spectra of $2<z<2.5$ quasars obtained at AAT.
The spectrum in each panel is represented by a {\it thin solid line}.
The model ({\it thick line}) is composed of a continuum ({\it dotted line}),
Fe {\sc ii} emission, H$\beta$, and [O {\sc iii}]
({\it long dashed lines}).}
\label{spectra_aat}
\end{figure*}

\begin{figure*}
\plotone{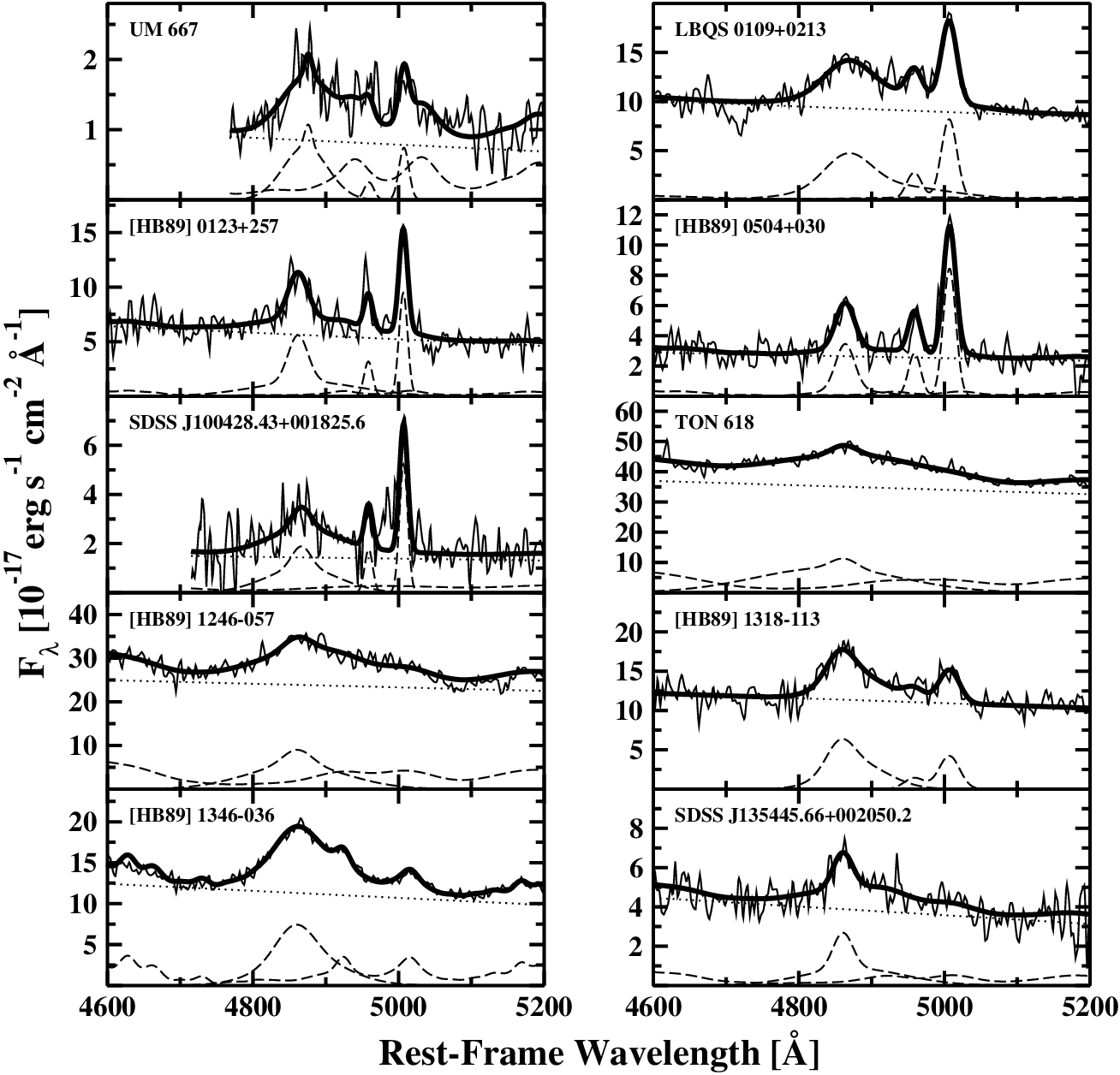}
\caption
{IR spectra of $2<z<3.5$ quasars obtained at TNG.
Symbols are as in Fig.~\ref{spectra_aat}.}
\label{spectra_tnga}
\end{figure*}

\begin{figure*}
\plotone{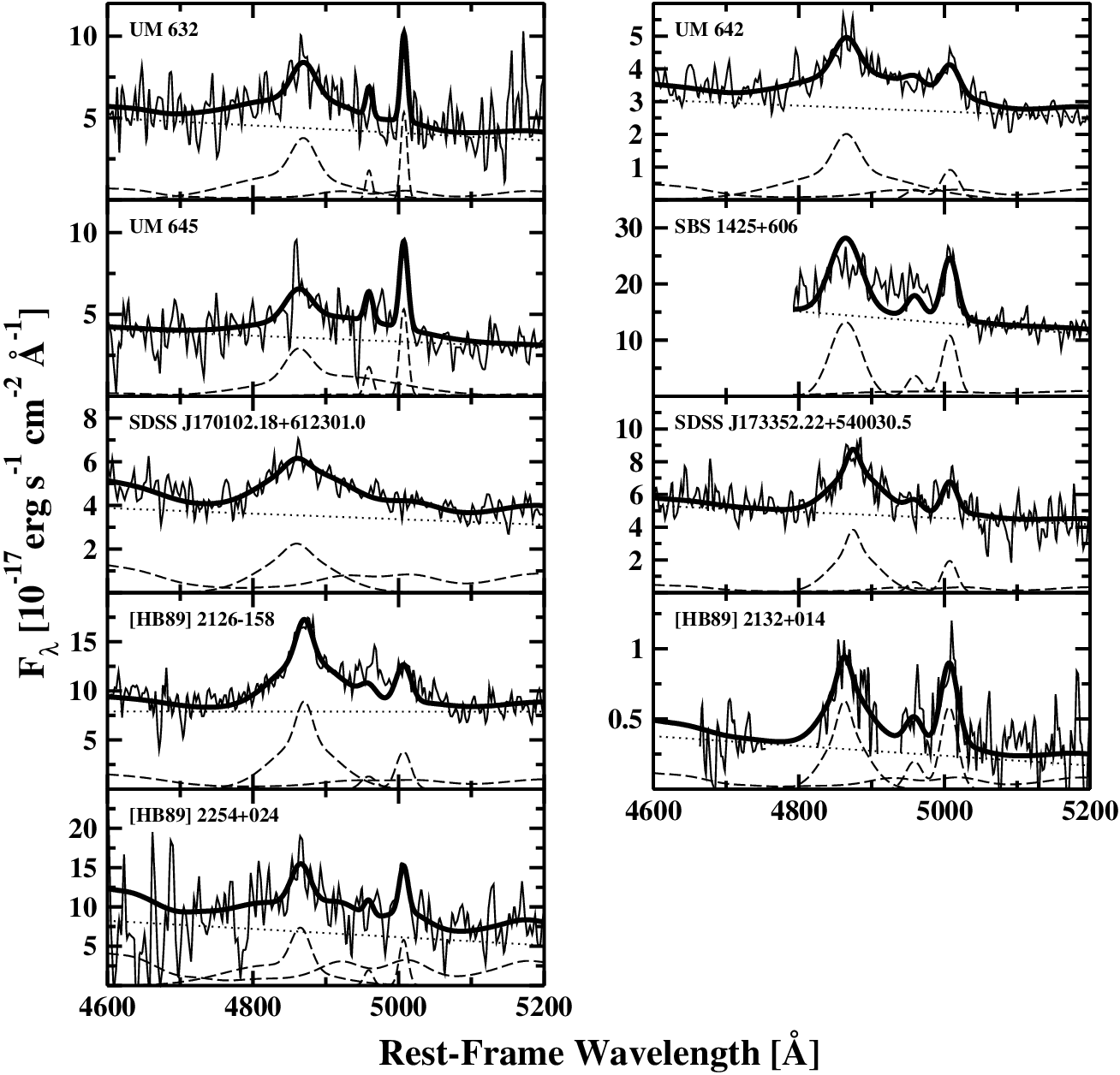}
\caption
{IR spectra of $2<z<3.5$ quasars obtained at TNG ({\it continued}).
Symbols are as in Fig.~\ref{spectra_tnga}.}
\label{spectra_tngb}
\end{figure*}

The two-dimensional spectra (quasars and standard stars) from both
observatories were reduced using standard {\sc IRAF}\footnote{IRAF (Image
Reduction and Analysis Facility) is distributed by the National Optical
Astronomy Observatories, which are operated by AURA, Inc., under cooperative
agreement with the National Science Foundation.} routines. All spectra were
pair-subtracted to remove most of the background, which is swamped by strong
telluric features, such as OH emission, leaving some residuals perpendicular
to the spatial axis. For each observatory we used a lamp spectrum as a
flat-field image to correct each spectrum for pixel-to-pixel variations. Each
spectrum was fitted by a low order polynomial, and a secondary background
subtraction, along the dispersion, was carried out. An extraction window of
$1\arcsec$ ($0.5\arcsec$) was used to extract each of the AAT (TNG) spectra,
respectively. Wavelength calibration was achieved using Xe and Ar arc lamps.
Finally, all spectra were rebinned to a common resolution of
$\sim600$~km~s$^{-1}$, matching the resolution of the TNG spectra taken
with the 1\arcsec\ slit.

The spectral shape of the quasars was recovered by dividing each
spectrum by a spectrum of a standard star, after the stellar
absorption features were carefully removed. The quasar/star ratio
was then multiplied by a stellar model in the form of a black body curve,
with a temperature corresponding to the star's tabulated spectral type.
Some of our standard stars have spectra that appear in the
library of near-IR stellar spectra of Lan\c{c}on \& Rocca-Volmerange (1992).
This enabled the removal of major intrinsic absorption features in our
standard stars, and also allowed to compare our stellar model to the archived
data. Flux calibration of the quasars was obtained by comparing
synthetic $H$ and $K$ magnitudes of the quasars (by integrating
counts across the dispersion) to the observed $H$ magnitudes
(see Table~\ref{log}), and using the flux density of Vega
as the zero-point flux.

Our flux calibration may be subject to relatively large uncertainties emerging
from different atmospheric absorption between a quasar and its corresponding
standard star. This uncertainty increases for the AAT spectra, as those are
cutoff short of the $H$ band blue end. For those spectra we integrated the
predicted counts after extrapolating the continuum (fitted by a linear
function) beyond $\sim1.55~\mu$m. We also used the b$_J$ magnitudes of the 2QZ
quasars to check for consistency with our derived rest-frame optical fluxes,
by converting those magnitudes to observed 4400\AA\ fluxes and extrapolating
them to 5100\AA\ assuming a fiducial quasar continuum function of the form:
$f_{\nu}\propto \nu^{-0.5}$. We found that our $H$ and b$_J$ magnitudes
(reflected to the observed $H$ band) were consistent to within $\pm0.25$
magnitudes. We note that those two magnitudes were measured at different
epochs, separated by $\sim20$ years in time, thus constraining the variability
amplitude of each quasar to $\sim25\%$ over a time scale of $\sim6$ yr in the
rest-frame (although there is a scope for variability within this long
period of time).
Luminosities were obtained by using the following cosmological parameters:
$\Omega_{\Lambda}=0.7$, $\Omega_{\rm m}=0.3$, and $H_{0}=70$ \kms\ Mpc$^{-1}$.

In addition to our 29 sources, we added to our sample two quasars,
LBQS 0256$-$0019 and LBQS 0302$-$0000, for which rest-frame
UV spectra were available from the Sloan Digital Sky Survey (SDSS;
Stoughton \et 2002) archive, and a high-quality rest-frame
optical spectrum including \hb\ was kindly provided by M. Dietrich (private
communication; see Dietrich \et 2002). As a result, our high redshift
sample consists of 31 luminous quasars.

Our new sample was complemented by 61 $z<2$ AGNs, from the original SN02
sample, all with published \hb\ data. Altogether this sample comprises 92
sources, including 10 NLS1s, with \ion{N}{5}, \ion{C}{4}, and \hb\ information.
One of our sources is a broad absorption line quasar ([HB89]~$1246-057$),
based on its UV spectrum, and 31 of our sources (8 at $2<z<3.5$, and the rest
at low-$z$ including one NLS1) were identified as being radio
loud quasars (RLQ), following the discrimination between RLQ and radio quiet
quasars (Kellerman \et 1989), and using optical and radio fluxes
from Veron-Cetty \& Veron (2003).

\section{Measurements and Data Analysis}
\label{analysis}
\subsection{\hb\ width}
\label{hbwidth}

The \hb\ region in the final spectra was fitted with a model
consisting of a linear continuum, the Boroson \& Green (1992) \ion{Fe}{2}
emission template, and a multi-Gaussian fit to the emission lines. We
also fitted the H$\alpha$ region in the TNG spectra (where present) in the
same way, but these data are left out of the current analysis.
In the first step we fitted a linear continuum and one Gaussian to
\hb\ to get a rough estimate of its full width at half maximum
(FWHM). The next step was to broaden the \ion{Fe}{2} emission template by
convolving it with a Gaussian whose width was identical to FWHM(\hb).
We then fitted the spectrum with a linear continuum, the broadened
\ion{Fe}{2} template, shifted in flux by a scaling factor, and multi-Gaussian
components for the emission lines. Two Gaussians with differing widths were
assigned to \hb\ and two Gaussians with identical widths were assigned to the
two [\ion{O}{3}]~$\lambda\lambda 4959,5007$ lines. The fluxes of the
[\ion{O}{3}] lines were constrained to the theoretical ratio
I([\ion{O}{3}]~$\lambda5007$)/I([\ion{O}{3}]~$\lambda4959$)$=2.95$,
and the widths of the two Gaussian components of \hb\ were free to vary.

The final spectra, best fit model, and model components, are plotted in
Figures~\ref{spectra_aat}--\ref{spectra_tngb}.
These fits are the ones used for FWHM(\hb) listed in Table~\ref{fit}.
The final FWHM of the entire \hb\ profile was measured, as well as the FWHM of
the [\ion{O}{3}] lines. We estimated the uncertainty on the width of the lines
by slightly shifting the parameters from their best-fit values and
marking the most extreme values that would still represent a
reasonable fit to the \hb\ complex. In almost all of the cases, those
uncertainties were smaller than the instrumental resolution ($\sim600$
\kms) for the TNG and AAT spectra, with UM~645, [HB89]~2254$+$024,
2QZ J002830.4$-$281706, and 2QZ J231456.8$-$280302 being the only exceptions
(see Figures~\ref{spectra_aat}--\ref{spectra_tngb}).

\begin{deluxetable*}{lccccccc}
\tablecolumns{8}
\tablewidth{0pt}
\tablecaption{Emission Line and Continuum Measurements
\label{fit}}

\tablehead
{
\colhead{Quasar Name} &
\colhead{Log} &
\multicolumn{2}{c}{FWHM(H$\beta$)} &
\colhead{Log} &
\colhead{$L/L_{\rm Edd}$} &
\colhead{N {\sc v}/C {\sc iv}} &
\colhead{Reference} \\
\colhead{} &
\colhead{$\lambda L_{\lambda}(5100)$} &
\colhead{Best Fit} &
\colhead{Direct} &
\colhead{$M_{\rm BH}$} &
\colhead{} &
\colhead{} &
\colhead{for} \\
\colhead{} &
\colhead{[erg s$^{-1}$]} &
\colhead{[\kms]} &
\colhead{[\kms]} &
\colhead{[\msun]} &
\colhead{} &
\colhead{} &
\colhead{UV data} \\
\colhead{(1)} &
\colhead{(2)} &
\colhead{(3)} &
\colhead{(4)} &
\colhead{(5)} &
\colhead{(6)} &
\colhead{(7)} &
\colhead{(8)}
}
\startdata
2QZ J001221.1$-$283630      & 46.26 & 1915 & 2382 & 8.71 & 1.64 & 0.84
& 1 \\
2QZ J002830.4$-$281706      & 46.58 & 4833 & 5295 & 9.71 & 0.35 & 0.47
& 1 \\
UM 667                      & 46.28 & 3135 & 2244 & 9.15 & 0.62 & 0.81 
& 2 \\
LBQS 0109$+$0213            & 46.80 & 5781 & 5241 &10.00 & 0.30 & 0.82
& 3 \\
\protect{[HB89]} 0123$+$257 & 46.57 & 2406 & 2763 & 9.09 & 1.38 & 0.77
& 2 \\
2QZ J023805.8$-$274337      & 46.57 & 3437 & 3691 & 9.40 & 0.68 & 0.90
& 1 \\
SDSS J024933.42$-$083454.4  & 46.38 & 5230 & 6601 & 9.66 & 0.25 & 1.07
& 4 \\
\protect{[HB89]} 0329$-$385 & 46.71 & 7035 & 8165 &10.11 & 0.18 & 0.86
& 5 \\
\protect{[HB89]} 0504$+$030 & 46.32 & 2046 & 2402 & 8.80 & 1.52 & 0.48
& 6 \\
SDSS J100428.43$+$001825.6  & 46.44 & 3442 & 3586 & 9.33 & 0.60 & 1.53 
& 7 \\
TON 618                     & 47.31 &10527 &14905 &10.82 & 0.14 & 1.60
& 8 \\
\protect{[HB89]} 1246$-$057 & 47.16 & 5817 & 6842 &10.22 & 0.41 & 1.82
& 5 \\
\protect{[HB89]} 1318$-$113 & 46.89 & 4150 & 4898 & 9.75 & 0.61 & 0.56
& 5 \\
\protect{[HB89]} 1346$-$036 & 46.88 & 5110 & 5129 & 9.94 & 0.41 & 0.78
& 5 \\
SDSS J135445.66$+$002050.2  & 46.49 & 2627 & 2208 & 9.13 & 1.08 & 2.93
& 7 \\
UM 629                      & 46.56 & 2621 & 2521 & 9.16 & 1.15 & 0.80
& 7 \\
UM 632                      & 46.54 & 3614 & 3828 & 9.43 & 0.60 & 0.28
& 6 \\
UM 642                      & 46.29 & 3925 & 5478 & 9.35 & 0.40 & 0.51
& 7 \\
UM 645                      & 46.31 & 3966 & 4972 & 9.37 & 0.40 & 0.29
& 7 \\
SBS 1425$+$606              & 47.38 & 3144 & 6769 & 9.82 & 1.71 & 0.47 
& 2 \\ 
SDSS J170102.18$+$612301.0  & 46.34 & 5760 & 5944 & 9.72 & 0.20 & 1.19
& 7 \\
SDSS J173352.22$+$540030.5  & 47.00 & 3078 & 5460 & 9.57 & 1.26 & 1.79  
& 7 \\
\protect{[HB89]} 2126$-$158 & 47.25 & 3078 & 3354 & 9.72 & 1.58 & 0.77 
& 2 \\
\protect{[HB89]} 2132$+$014 & 45.77 & 2505 & 2801 & 8.65 & 0.61 & 0.80 
& 2 \\
2QZ J221814.4$-$300306      & 46.54 & 2986 & 3070 & 9.27 & 0.88 & 0.96
& 1 \\
2QZ J222006.7$-$280324      & 47.22 & 5238 & 5660 &10.16 & 0.53 & 1.67
& 1 \\
\protect{[HB89]} 2254$+$024 & 46.45 & 2597 & 1954 & 9.09 & 1.07 & 0.59
& 6 \\
2QZ J231456.8$-$280102      & 46.31 & 3459 & 2011 & 9.26 & 0.53 & 1.21
& 1 \\
2QZ J234510.3$-$293155      & 46.32 & 3908 & 3453 & 9.37 & 0.42 & 1.59
& 1 \\
\enddata
\tablerefs{
1. 2QZ archive (Croom \et 2004).
2. Dietrich \& Wilhelm-Erkens (2000).
3. Forster \et (2001).
4. SDSS first data release (Abazajian \et 2003).
5. Osmer \& Smith (1977).
6. Baldwin, Wampler, \& Gaskell (1989).
7. SDSS early data release (Stoughton \et 2002). 
8. Baldwin \& Netzer (1978).
}
\end{deluxetable*}

We have also obtained direct measures of the fluxes and FWHMs of
\hb\ and [\ion{O}{3}], by integrating across the \ion{Fe}{2}-subtracted
features, for the flux measurements, and by taking the width at half the peak
value for each \ion{Fe}{2}-subtracted feature, in the FWHM measurements.
Both types of measurements, the best-fit value, and the 'direct' feature
method, appear in Table~\ref{fit}, and in Paper II. The difference between
both methods can be regarded as an estimate of the uncertainty on those
measurements, although these are not pure statistical errors. This paper
presents the \hb\ line width for our sources. Other line properties, and in
particular those of [\ion{O}{3}] and \ion{Fe}{2}, are discussed in Paper II.

It is expected that each \hb\ profile in our $2<z<3.5$ sample is a blend of
a narrow component, emitted from the narrow-line region (NLR), and a broad
BLR component, that we will use to determine \mbh. Resolving each of
those two physical components can be done if there is an obvious NLR spike
in the \hb\ profile, as is the case for several broad-line Seyfert 1 galaxies
(BLS1s).
In cases where a narrow spike is undetected, the NLR contribution to the \hb\
emission can in principal be estimated by taking
FWHM([\ion{O}{3}]~$\lambda5007$) as its
width, and its flux as a certain fraction of the [\ion{O}{3}]~$\lambda5007$
flux. This method is highly model-dependent, and in many of our sources we
do not detect [\ion{O}{3}] emission at all (see Paper II). The two Gaussians
we fit to \hb\ do not attempt to recover the {\em physical} NLR and BLR
components, but merely to represent a fit to the entire profile. In luminous
AGNs the NLR-to-BLR flux ratio, at least in the UV lines, is of the order of a
few percent (e.g., Wills \et 1993), resulting in a negligible NLR
contribution to the line flux. Since we also do not detect a significant NLR
spike {\em in any} of our \hb\ profiles, we assume that the observed line
represents entirely the BLR contribution.

\begin{figure}
\epsscale{0.9}
\plotone{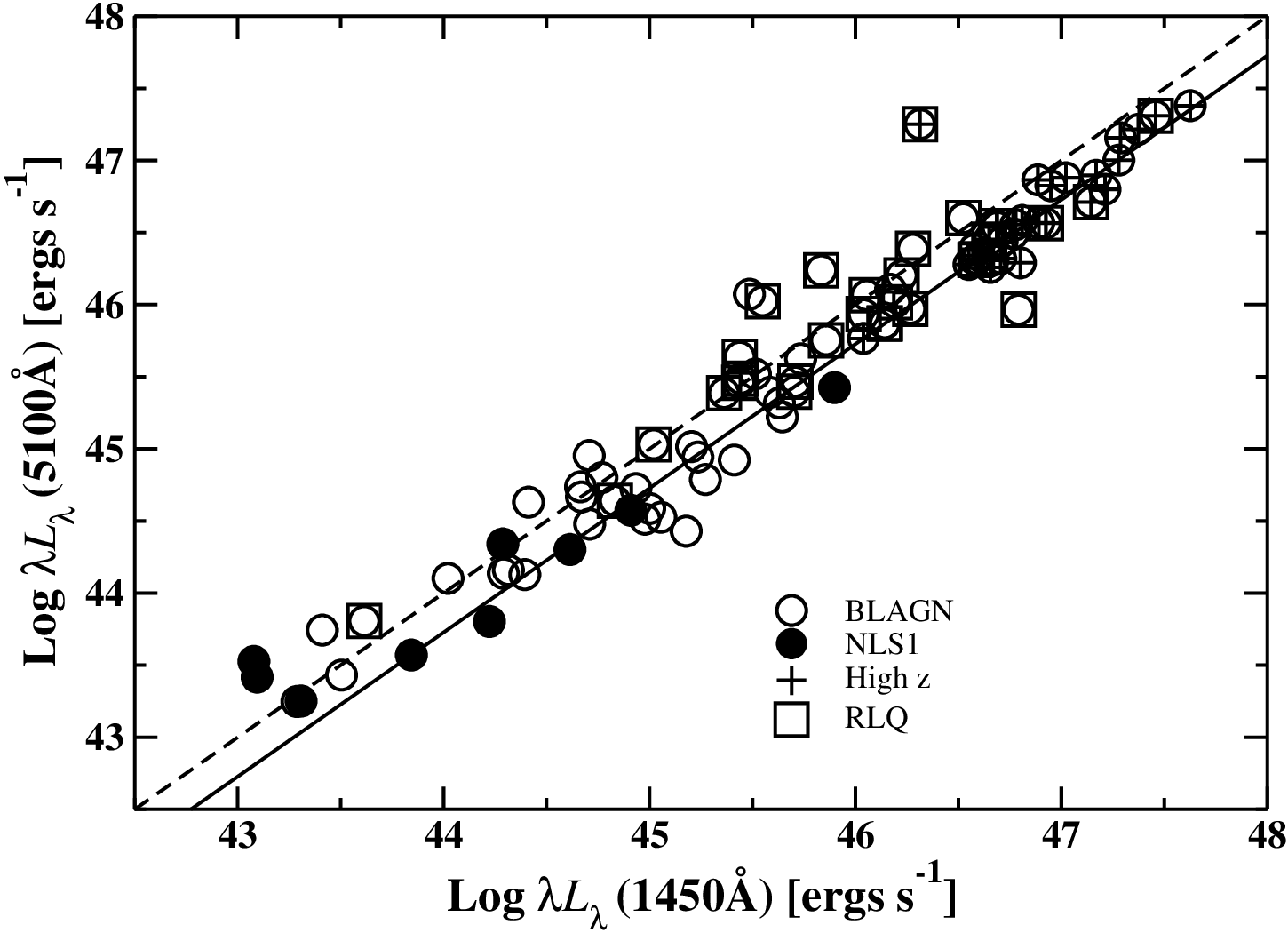}
\caption
{Monochromatic luminosities at 5100\AA\ vs. 1450\AA\ for the sample of 92 AGNs.
Empty circles are broad-line AGNs (BLAGNs) with FWHM(H$\beta$)$>$1500 km
s$^{-1}$, and {\it plus signs} indicate $2<z<3.5$ quasars.
NLS1s (with FWHM(H$\beta$)$<$1500 km s$^{-1}$; see SN02) are marked with
({\it filled circles}), and RLQs are marked with {\it empty squares}.
The {\it dashed line} marks a 1:1 correspondence, and the {\it solid line}
marks an empirical ratio of
$\lambda L_{\lambda}(1450)/\lambda L_{\lambda}(5100)=\sqrt{(5100/1450)}$
assuming $f_{\nu}\propto \nu^{-\alpha}$ with $\alpha=0.5$. Most AGNs lie closer
to the dashed line, indicating somewhat larger $\alpha$.}
\label{L5100vs1450}
\end{figure}

\begin{figure*}
\epsscale{1.0}
\plotone{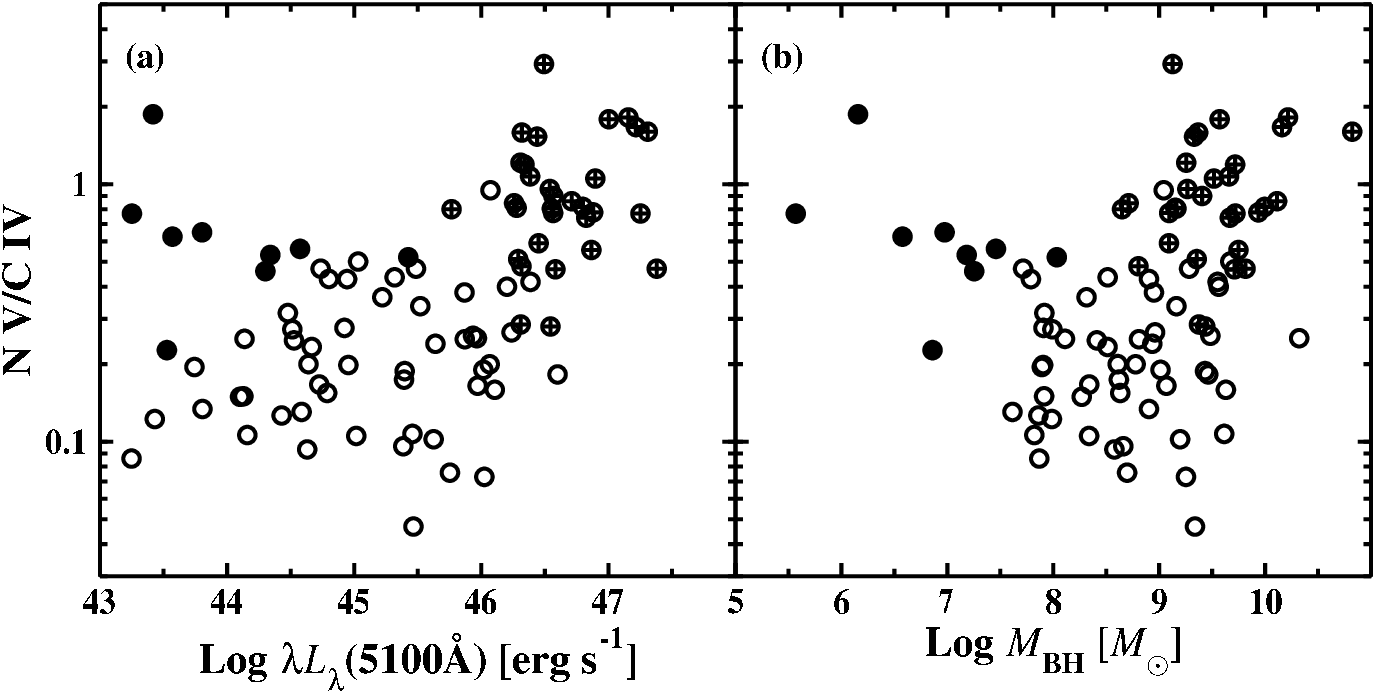}
\caption
{N {\sc v}/C {\sc iv} vs. ({\it a}) luminosity, and ({\it b}) \mbh. Symbols are
as in Fig.~\ref{L5100vs1450}. NLS1s clearly lie away from the trend followed by
BLAGNs and high-$z$ quasars.}
\label{diagrams_ab}
\end{figure*}

We have also obtained another measure of line-width, based on the entire
emission line flux, regardless of the profile. This is the inter-percentile
velocity (IPV) width, which we have used
for measuring the difference between wavelengths corresponding to 25\%
and 75\% of the integrated emission line flux across the entire line
profile, starting with the continuum on the blue side and ending with
the continuum on the red side of the line. Both the FWHM and IPV
methods have their own merits, although it is still not clear which
gives the closest estimate of the BLR velocity width (see Corbett \et 2003 and
references therein).
In most cases, the IPV widths of \hb\ are somewhat narrower than the
corresponding FWHM (by $\sim15$\%; for a Gaussian profile the IPV width
should be $\sim43$\% narrower than the corresponding FWHM), although both
methods give consistent line-widths within the uncertainties discussed above.

\subsection{Luminosity, BH mass, and accretion rate}
\label{ZLME}

We have derived monochromatic luminosities for our entire AGN sample
at the rest-frame 5100\AA\ band using our new IR observations, and data from
the literature, and compared them with the monochromatic luminosities at the
rest-frame 1450\AA\ band obtained from the literature, and from the archives
of the SDSS, and the 2dF quasar redshift survey (2QZ; Croom \et 2004). For
ten sources we could not find 5100\AA\ fluxes, including six sources from our
high-$z$ sample (see Table~\ref{log}). In those cases we obtained the
5100\AA\ luminosity by extrapolating from 1450\AA \ fluxes and assuming
(as explained in \S~\ref{observations}) a continuum function of the form:
$f_{\nu}\propto \nu^{-0.5}$. Figure~\ref{L5100vs1450} shows this comparison,
along with two lines, indicating a 1:1 correspondence, and a
1:$\sqrt{5100/1450}$ representing the expected ratio between the 1450\AA \ and
the 5100\AA \ bands for a $f_{\nu}\propto \nu^{-\alpha}$ continuum with
$\alpha=0.5$.
Most AGNs lie in between the two lines indicating perhaps a somewhat redder
($\alpha>0.5$) continuum for our sample. We also note that most RLQs in
our sample are slightly redder than the radio quiet objects, consistent with
the steep continuum slopes in RLQs found by Netzer \et (1995).
Since the luminosities in both bands are highly correlated, and in
order to avoid complications due to intrinsic reddening, we prefer to use
the 5100\AA\ luminosities\footnote{For a few nearby Seyfert 1 galaxies in
our sample, this choice might add contamination from the host galaxy.
However, in most cases such a contribution is much less than 50\%,
which is an acceptable uncertainty level, given the typical large
expected continuum variations.} over the 1450\AA \ luminosities in
determining BH masses and accretion rates in our entire sample.

Black hole masses and accretion rates were determined by the single-epoch mass 
determination method (Vestergaard 2002), which is based on the Kaspi \et
(2000, hereafter K00) $R_{\rm BLR} - \lambda L_{\lambda}(5100)$ correlation, 
where $\lambda L_{\lambda}(5100)$ is the monochromatic luminosity at 5100 \AA.
In this method the BLR size is estimated from the measured
$\lambda L_{\lambda}(5100)$ and the mass estimates follows from assuming
Keplerian motion of the \hb\ emitting gas.
The scaling used here is based on a re-analysis of the K00 measurements
(Peterson \et 2004; Kaspi et al., in preparation)
that gives
$R_{\rm BLR} \propto \lambda L_{\lambda}(5100)^{0.6}$ with a somewhat different
size scaling factor.
It also uses a somewhat different mass scaling factor that multiplies the
FWHM(\hb) term in the mass determination equation (see K00). The result,
\begin{equation}
\label{eq:def_mass}
M_{\rm BH}=6.2 \times 10^6 \left [ \frac{\lambda
L_{\lambda}(5100)}{10^{44} \
{\rm erg \ s^{-1}}} \right ] ^{0.6} \left [ \frac{{\rm
FWHM}({\rm H}\beta)}{10^3 \ {\rm km \ s^{-1}}} \right ] ^2 \ M_{\sun} ,
\end{equation}
gives slightly larger masses compared with K00 and Netzer (2003)
\footnote {In fact, the most recent result of Kaspi et al., in prep., shows
that the mass scaling factor used here is $\sim25$\% larger than theirs,
but both factors are consistent within the uncertainties.}.
We then convert the observed luminosity and the deduced masses to accretion
rates, represented by $L_{\rm Bol}/L_{\rm Edd}$, by assuming
$L_{\rm Bol} = 7 \times \lambda L_{\lambda}(5100)$. 
The factor 7 is in the middle of the range (5--9) representing observed and
theoretically deduced spectral energy distributions (SEDs; see Netzer 2003).
The result is
\begin{equation}
\label{eq:def_eddington2}
L_{\rm Bol}/L_{\rm Edd} = 0.75 \left [ \frac{\lambda
L_{\lambda}(5100)}{10^{44}
\ {\rm erg \ s^{-1}}} \right ] ^{0.4} \left [ \frac{{\rm
FWHM}({\rm H}\beta)}{10^3 \ {\rm km \ s^{-1}}} \right ] ^ {-2} \, .
\end{equation}
Table~\ref{fit} lists luminosity, \mbh, and $L_{\rm Bol}/L_{\rm Edd}$
(hereafter $L/L_{\rm Edd}$) for our new $2<z<3.5$ quasars. By combining the
measurement uncertainties on FWHM(\hb), as given in Table~\ref{fit} and
discussed in \S~\ref{hbwidth}, with the small ($\sim25$\%) uncertainty on the
luminosity, we obtain typical uncertainties on \mbh\ and on $L/L_{\rm Edd}$,
which are not larger than a factor of two, and therefore do not affect our
main results as outlined below.

A similar determination of \mbh, and $L/L_{\rm Edd}$
was applied to LBQS 0256$-$0019, LBQS 0302$-$0000, and 61 sources from the
SN02 sample. It is worth noting at this point that, within our new (albeit
small) sample of luminous high-$z$ quasars, there is no clear relation between
\mbh\ and radio loudness (see Tables~\ref{log}~\& ~\ref{fit}), as was
previously suggested in several studies, e.g., Franceschini, Vercellone,
\& Fabian (1998), and it is therefore consistent with the results of
Woo \& Urry (2002).

\begin{figure*}
\epsscale{0.7}
\plotone{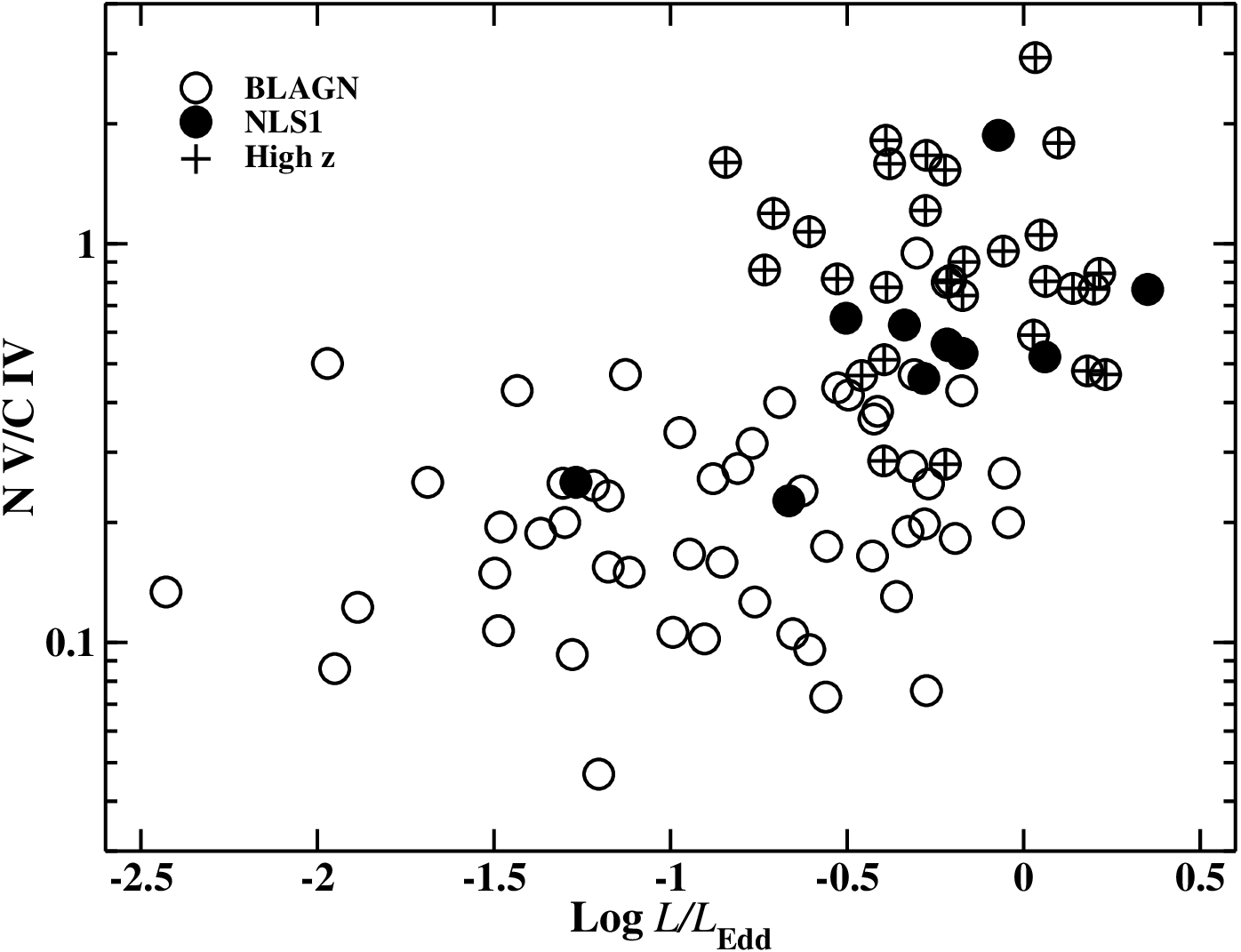}
\caption
{N {\sc v}/C {\sc iv} vs. accretion rate. Symbols are similar to those
in Fig.~\ref{L5100vs1450}.
A strong metallicity--accretion rate correlation {\em for all AGNs}
is apparent, and most NLS1s are found in the same region of parameter
space that is shared by the high-$z$ quasars. Note that the location
of Mrk~766 (the NLS1 with the lowest accretion rate in this diagram)
may be affected by strong intrinsic reddening (see text).}
\label{diagrams_c}
\end{figure*}

\subsection{BLR metallicity}
\label{Z}

Following HF93, we use \nvciv\ as the BLR metallicity indicator for our
sample. This line ratio is easier to measure than most other emission
lines in a typical spectrum, and was considered by Hamann \et (2002) as a
robust indicator since it minimizes non-abundance effects, such as
temperature, and SED.
About half of our quasars have rest-frame UV data available in the literature,
and we complemented these data with measurements of \ion{N}{5} and \ion{C}{4}
for the rest of the sample, which are mostly recently detected quasars.
Rest-frame UV spectra of those sources were obtained from the SDSS and
2QZ archives. The standard SDSS data allows reasonably accurate flux
calibration and SED determination.
The shape of the non-flux calibrated 2QZ spectra is more problematic and
was obtained by dividing each spectrum by a quadratic, which provides
a reasonable estimate to the spectral response function of the 2dF
(Folkes \et 1999).
Each SDSS and 2QZ spectrum was fitted with a power-law continuum, and
two Gaussians for the \ion{C}{4} emission line. The \ion{C}{4} profile
was then used as a template for all other rest-frame UV emission
lines, and in particular \ion{N}{5}. We list \nvciv measurements for
our 29 $2<z<3.5$ quasars in Table~\ref{fit}. The uncertainties on the
measurement of this ratio vary between $\sim10$\% for the high-quality
spectra and $\sim30$\% for the poorer quality ones (see also SN02).

\section{Results}
\label{results}
\subsection{A new metallicity--accretion rate relationship in AGNs}
\label{ZE}

In Figure~\ref{diagrams_ab} we plot \nvciv\ vs.
({\it a}) $\lambda L_{\lambda}(5100)$, and ({\it b}) \mbh\
for our sample of 92 AGN. One can see that the $Z-L$, and $Z-M_{\rm BH}$
relations show considerable scatter, mainly due to the NLS1s which have low
BH masses and luminosities. The first of those relations has been shown
and discussed by SN02 for a slightly larger sample. The scatter is
significantly reduced once metallicity is plotted against accretion rate as
shown in Fig.~\ref{diagrams_c}. In this case, {\em all} AGNs follow a strong
metallicity--accretion rate (\nvciv\ --$L/L_{\rm Edd}$;
hereafter $Z-L/L_{\rm Edd}$) correlation with NLS1s {\em being no exception}.

Table~\ref{spearman} lists the Spearman rank correlation coefficients
($r_{\rm S}$) and the chance probabilities ($P$) for each of the
three relations. We also tested the correlations by removing the 10
NLS1s from the sample. As can be seen from Table~\ref{spearman}, this increases
$r_{\rm S}$ considerably in the $Z-L$ and $Z-M_{\rm BH}$ cases, but their
removal from the $Z-L/L_{\rm Edd}$ diagram leaves $r_{\rm S}$ almost unchanged.
We note that formally, both the $Z-L$ and the $Z-M_{\rm BH}$ correlations
are significant ($P\ll0.01$). This represents the fact that our sample contains
only 10 NLS1s, out of 92 sources, (representing well their fraction in the AGN
population). However, this sub-group clearly lies away from the general trend
in Figures~\ref{diagrams_ab}{\it a} \& \ref{diagrams_ab}{\it b} emphasizing
the fact that they do not share the same properties with other low
luminosity/\mbh\ AGNs.

\begin{deluxetable}{lccl}
\tablewidth{0pt}
\tablecaption{Spearman Rank Correlation Coefficients
\label{spearman}}

\tablehead
{
\colhead{N {\sc v}/C {\sc iv}} &
\colhead{Number of} &
\colhead{$r_{\rm S}$} &
\colhead{$P$} \\
\colhead{vs.} &
\colhead{Sources} &
\colhead{} &
\colhead{} \\
\colhead{(1)} &
\colhead{(2)} &
\colhead{(3)} &
\colhead{(4)}
}
\startdata
$L/L_{\rm Edd}$\tablenotemark{a} & 92 & 0.57 & $<10^{-5}$ \\
$L/L_{\rm Edd}$\tablenotemark{a} & 82\tablenotemark{b} & 0.54 &
$<10^{-5}$ \\
$M_{\rm BH}$\tablenotemark{a}    & 92 & 0.35 & $3.1\times 10^{-4}$ \\
$M_{\rm BH}$\tablenotemark{a}    & 82\tablenotemark{b} & 0.53 &
$<10^{-5}$ \\
$\lambda L_{\lambda}(5100)$ & 92 & 0.53 &
$<10^{-5}$ \\
$\lambda L_{\lambda}(5100)$ & 82\tablenotemark{b} & 0.71 &
$<10^{-5}$ \\
$L/L_{\rm Edd}$\tablenotemark{c} & 82 & 0.32 &
$1.6 \times 10^{-3}$ \\
$L/L_{\rm Edd}$\tablenotemark{c} & 72\tablenotemark{b} & 0.41
& $1.7 \times 10^{-4}$ \\
\enddata
\tablenotetext{a}{Calculated from $\lambda L_{\lambda}(5100)$ and
FWHM(\hb)}
\tablenotetext{b}{NLS1s excluded}
\tablenotetext{c}{Calculated from $\lambda L_{\lambda}(1450)$ and
FWHM(C {\sc iv})}
\end{deluxetable}

To summarize, our results indicate that the \nvciv\ line ratio is correlated
primarily with $L/L_{\rm Edd}$ and not with luminosity or BH mass.
It is the combination of luminosity {\em and} line-width (i.e., the accretion
rate) that makes all the AGNs in our sample, including NLS1s, follow the same
trend. We caution that our entire sample of 92 AGNs is neither complete nor
fully representative of the AGN population as a whole, and that our main
result may be subject to selection biases over the wide range of AGN
properties, that were not fully explored in this work.

\begin{figure}
\epsscale{1.1}
\plotone{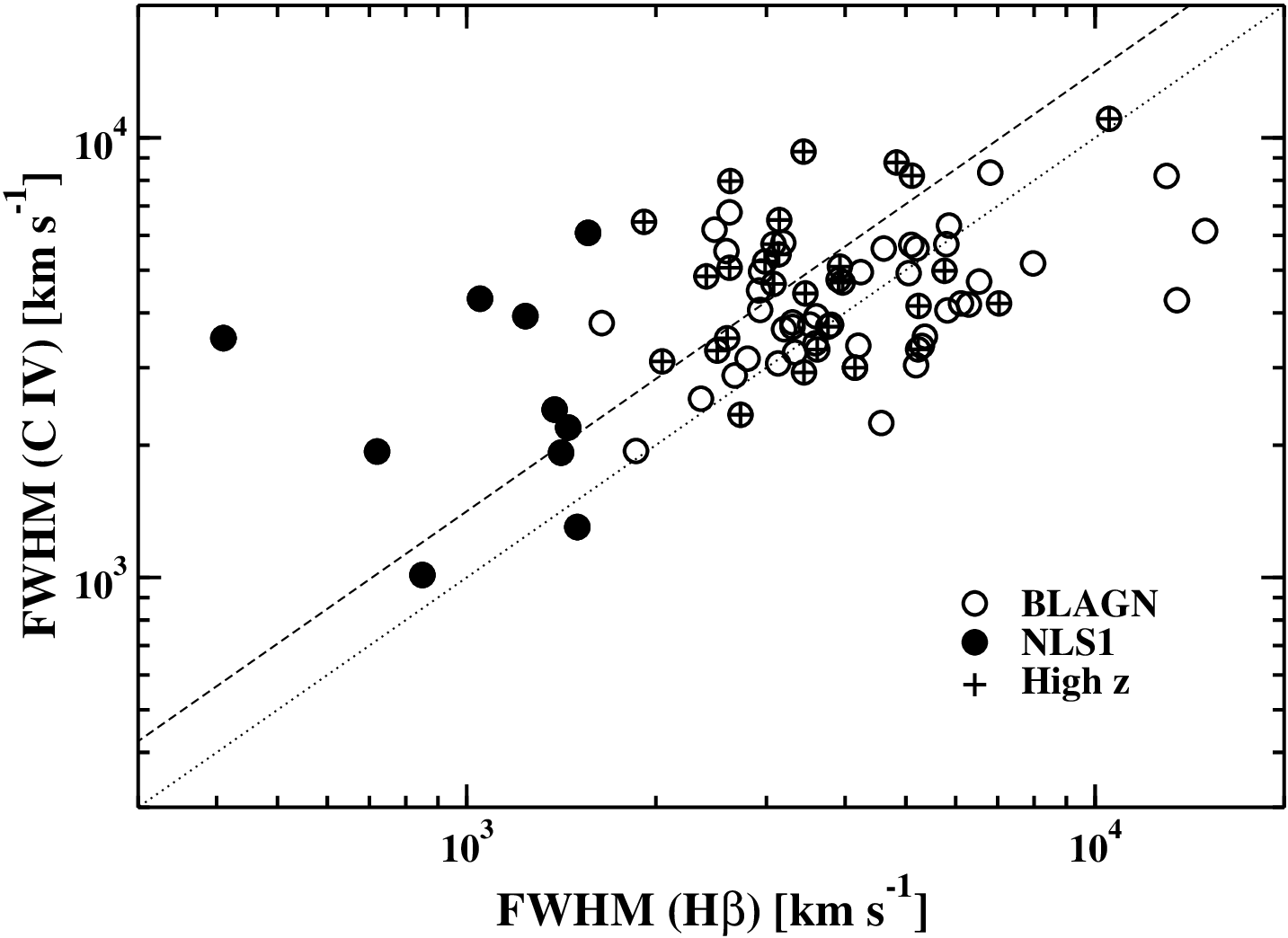}
\caption
{FWHM(C {\sc iv}) vs. FWHM(\hb) for 82 objects in our entire
sample. Symbols are similar to those in Fig.~\ref{diagrams_c}. A {\it dotted}
({\it dashed}) line represents a 1:1 (1:$\sqrt2$) relation, respectively.}
\label{hbc4}
\end{figure}

\subsection{FWHM(\hb) vs. FWHM(\ion{C}{4}) as accretion rate indicators}
\label{hbvsc4}

Reverberation BH masses were so far determined for 35 AGNs using \hb\
and the monochromatic luminosity at 5100\AA\ (K00, Peterson \et 2004),
to set the empirical $R-L$ relation on which all \mbh\ estimates are made.
For three AGNs, reverberation masses were also determined using \ion{C}{4}
(Peterson \& Wandel 2000), and were consistent with the masses obtained from
\hb. Vestergaard (2002) suggested that \ion{C}{4} can be used as a 'proxy'
to \hb\ and replace it as the reverberation \mbh\ indicator. It
is expected that the BLR components of both lines will satisfy
FWHM(\ion{C}{4})/FWHM(\hb)~$\simeq\sqrt2$, based on reverberation studies
of both lines that show a smaller mean emission distance for \ion{C}{4}
(Peterson \& Wandel 2000).

We have measurements of both \hb\ and \ion{C}{4} in 82 AGNs of our sample,
and the width of one line vs. the width of the other is plotted
in Fig.~\ref{hbc4}. One can
clearly see that the widths of both lines do not follow a 1:1 correspondence
nor a 1:$\sqrt2$ dependence. In particular, NLS1s with narrow \hb\
lines show FWHM(\ion{C}{4}) which is not different than those of
'normal' BLS1s, a fact that was first noted by Rodriguez-Pascual \et (1997).
Broad \ion{C}{4} in NLS1s, and in other AGNs, has been attributed to high
velocity gas outflowing from the center rendering a broad and blueshifted tail
to this line (e.g., Leighly 2000).

To demonstrate the influence of
FWHM(\ion{C}{4}) on NLS1 \mbh\ estimates, we compared the Vestergaard (2002)
single-epoch \mbh\ estimated from the UV spectrum with the reverberation
\mbh\ from K00, directly measured from the \hb\ width, for four NLS1s. We find
that, on average, the \mbh\ UV estimates are larger by a factor of $\sim3$
than the \mbh\ measured by optical reverberation mapping. Hence, \mbh\
calculated from \ion{C}{4}, at least for NLS1s, may be systematically higher
than \mbh\ estimates using \hb. An opposite behavior is also apparent
in Fig.~\ref{hbc4}, namely narrow \ion{C}{4} for broad \hb.

\begin{figure}
\epsscale{1.0}
\plotone{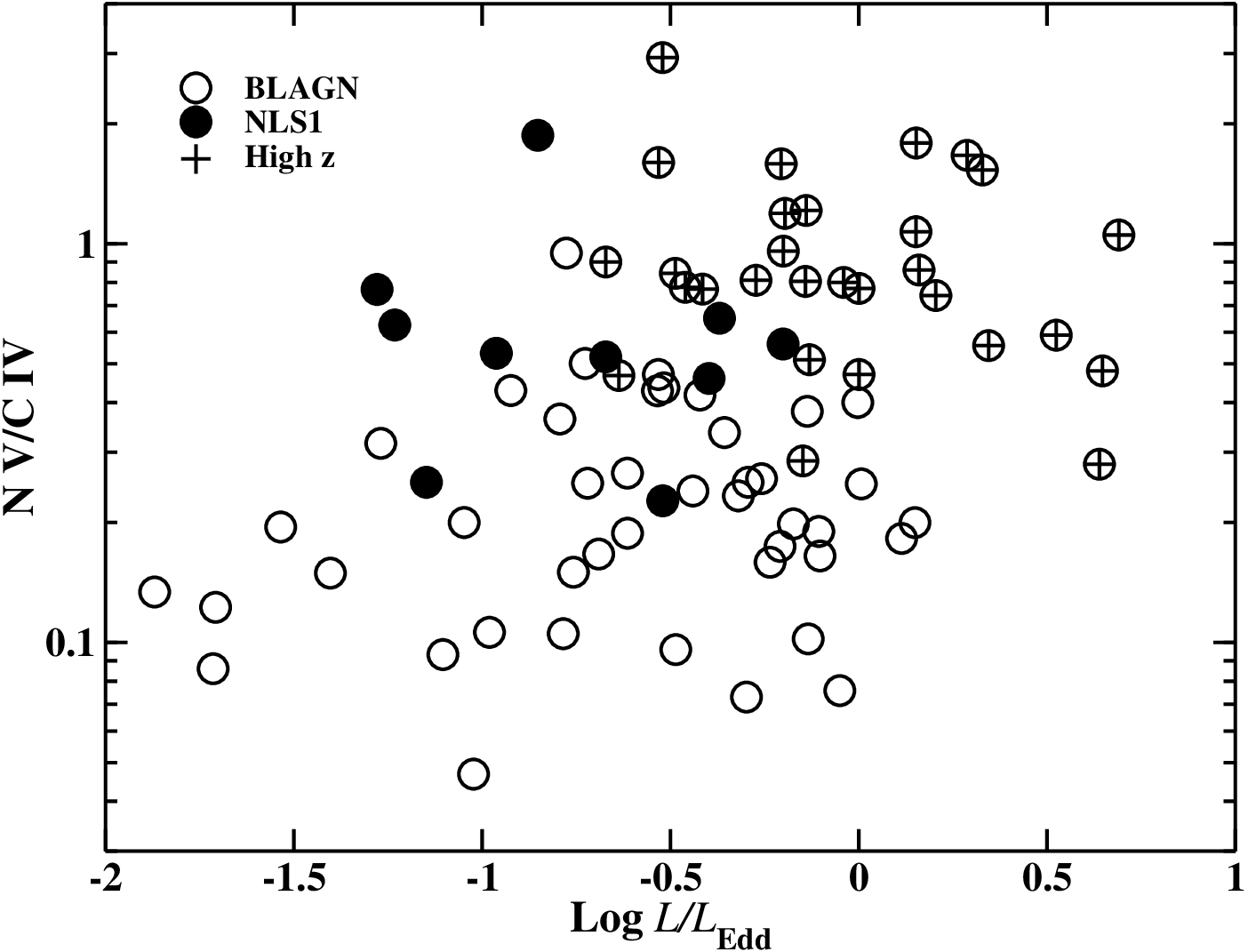}
\caption
{N {\sc v}/C {\sc iv} vs. accretion rate determined from a
combination of $\lambda L_{\lambda}(1450)$ and FWHM(C {\sc iv}).
Symbols are similar to those in Fig.~\ref{diagrams_c}.
Note the considerable scatter, compared with Fig.~\ref{diagrams_c},
as a consequence of using C {\sc iv} as an accretion rate indicator.}
\label{diagramsc4}
\end{figure}

To further illustrate the problematic use of \ion{C}{4} for individual sources,
we recalculated the BH mass and accretion rate, this time using the width of
\ion{C}{4}, together with the Vestergaard (2002) $R_{\rm BLR}-L$ relation.
Figure~\ref{diagramsc4} plots \nvciv\ vs. accretion rate,
and Table~\ref{spearman} lists the corresponding Spearman rank correlation
coefficients.
By inspection of Fig.~\ref{diagramsc4} and Table~\ref{spearman} it is
apparent that the correlation between the accretion rate calculated from the
\ion{C}{4} width, and \nvciv, is much weaker compared with the
accretion rate calculated from \hb, and that the $Z-L/L_{\rm Edd}$ now
has much more scatter. This result may also explain the lack of a
$Z-L/L_{\rm Edd}$ correlation in Warner, Hamann, \& Dietrich (2004). 

\section{Discussion}
\label{discussion}

The main result presented in this paper is a strong correlation
between metallicity (\'{a} la HF93) and accretion rate in a sample
of 92 AGNs spanning a broad range of AGN luminosity ($\sim10^{42}-10^{48}$
erg s$^{-1}$), line-width, and radio loudness. 
We also confirm the earlier SN02 finding that AGN samples that include
NLS1s do not show a clear $Z-L$ correlation.
The new relationship, although supported by a strong correlation, is not
free of scatter, and the effects of other physical properties are still
unknown. In this section we discuss potential
sources for intrinsic scatter in the $Z-L/L_{\rm Edd}$ diagram as well as
potential caveats in the derivation of BLR metallicity and
accretion rate. We also comment on the potential implications of our new
result on metal enrichment and BH growth scenarios in galactic nuclei,
and on the starburst--AGN connection.

\subsection{How accurate is the BH mass determination at high luminosity?}
\label{bhmass}

Our estimate of BH mass, and hence accretion rate, rely on the K00 relationship
applied to a sample of 34 AGNs with BH masses measured via reverberation
mapping and modified, slightly, in Kaspi et al., in preparation.
That relationship is limited by the small number of sources, the luminosity
range ($\sim10^{42}-10^{45}$), the assumed constant SED, and perhaps other
AGN properties. Mass estimates for sources beyond this luminosity range
are therefore uncertain. Such extrapolations lead to BH masses
exceeding $10^{10}$\msun\ in this work and in Netzer (2003).
Hence, if we have somewhat overestimated \mbh\ in our luminous
AGN sample, then the accretion rate must have been underestimated.
This will move luminous quasars to the right hand side of
Fig.~\ref{diagrams_c} and will produce an even stronger
$Z-L/L_{\rm Edd}$ correlation. However, a very different $L-M_{\rm BH}$
relationship at high luminosity may introduce more scatter into some
of the correlations presented here.

\subsection{Is \nvciv\ an adequate metallicity indicator for all AGNs?}
\label{adequate}

Using a grid of photoionization calculations, performed with the locally
optimally emitting cloud model (LOC; e.g., Baldwin \et 1995), Hamann
\et (2002) found that \nvciv\ can trace the BLR gas metallicity, and
it only weakly depends on the chosen SED and density. Our new results show
that \nvciv\ depends primarily on $L/L_{\rm Edd}$, but it is not yet clear
whether high \nvciv\ is indeed indicative of high metallicities.
In particular, it is crucial to compare the observations of this line ratio
to those of the model-independent weak-lines proposed by Shields (1976).
Our investigation of this issue is beyond the scope of this work
(see also Dietrich \et 2003).

Part of the scatter in the $Z-L/L_{\rm Edd}$ diagram may be contributed by
the yet unknown effect of BLR density on the way BLR metallicity is derived.
The problem has been mentioned by HF93, but the observational verification
is still missing. It is reasonable to assume that high metal contents are
associated with high densities in the nuclear region which are indicators of
recent star formation. In particular, NLS1s which have high ratios of \nvciv,
are also known to have high BLR densities, derived from the
\ion{Si}{3}]~$\lambda1892$/\ion{C}{3}]~$\lambda1909$
density-sensitive line ratio (e.g., Wills \et 1999). We have measured this
line ratio in our sample (available for about half of our sources), and found
no clear trend with either metallicity or accretion rate. The two emission
lines used for this ratio, which are fairly close to each other, are hard to
measure and to deblend. Therefore the potential effect of the BLR density
remains unclear at this stage.

The continuum SED may also affect the way \nvciv\ depends on BLR metallicity.
It is not yet clear whether the SED depends on luminosity, or perhaps the
accretion rate, but if the SED in high accretion rate sources is
significantly different from the ones used by Hamann \et (2002) to calculate
\nvciv, then it may cause some deviation from the canonical $Z-L$ dependence.
In particular, NLS1s appear to have SEDs which are somewhat different
from 'standard' AGN SEDs (e.g., Romano \et 2004), and related perhaps to the
putative (hotter) slim accretion disk (e.g., Wang \& Netzer 2003).
As explained in \S~\ref{ZLME}, a different SED may also affect our derived
$L/L_{\rm Edd}$.

A viable model of AGN evolution is that they start off with
high accretion rates, when fuel supply is abundant, and gradually
move towards lower accretion rates as the fuel supply dwindles.
If the metal content of the gas does not change as a consequence
of the fuel shortage, one can expect an AGN to move horizontally, from right
to left, across the $Z-L/L_{\rm Edd}$ diagram (Fig.~\ref{diagrams_c}),
as a function of time.
In other words, similar metallicities may be observed for a range
of accretion rates. The dependence on age may be further complicated by
the AGN duty cycle (assumed to be $\sim10^{7}-10^{8}$ yr), as we do not know
the number of cycles each AGN in our sample has gone through. If
metal enrichment is strongly related to AGN fueling, then a number
of active episodes may contribute to metal enrichment provided
the metal rich gas is not being depleted or blown out of the nucleus.
This can lead to a $Z-L/L_{\rm Edd}$ dependence which is more complicated
than assumed here.

\subsection{Intrinsic reddening}
\label{red}

Intrinsic reddening may affect the location of an AGN on the $Z-L/L_{\rm Edd}$
diagram, by moving it diagonally towards the lower left corner in
Fig.~\ref{diagrams_c}, due to weaker \ion{N}{5} with respect to
\ion{C}{4} and lower luminosity. Correcting for intrinsic reddening is
problematic, and requires a broader spectral range to compare suitable line
ratios (e.g., Crenshaw \et 2002).
We did not attempt to correct the data for intrinsic reddening but
note that in the case of NLS1s, such a correction may be necessary,
since these objects are suspected to possess large amounts of dust and
gas (e.g., Crenshaw, Kraemer, \& Gabel 2003). We note that at
least one NLS1 in our sample (Mrk~766) may be strongly affected by intrinsic
reddening, which might explain its location in Fig.~\ref{diagrams_c},
away from the rest of the NLS1 population.

\subsection{Metal enrichment and BH growth in galactic centers}
\label{BHgrowth}

If \nvciv\ is an adequate BLR metallicity indicator, then our
$Z-L/L_{\rm Edd}$ correlation may have important implications for the
understanding of BH growth in galactic nuclei as well as for the
starburst--AGN connection. We find that luminous quasars at high-redshift,
as well as NLS1s in the nearby Universe, may be experiencing (or have
recently experienced) a vigorous star-forming episode, and their BHs
are in the stage of rapid growth. In this sense, the
NLS1s in our sample may be in the early stages of their (current)
activity, i.e., relatively 'young' systems, as suggested by Mathur (2000).
All the high-$z$ quasars in our sample also exhibit high accretion rates
and high metallicities and may be in a similar evolutionary stage.
These sources are thus the high luminosity analogs of NLS1s and should,
perhaps, be referred to as narrow-line type 1 quasars (NLQ1s).
However, our sample does not contain faint quasars at large distances and those
must be added to the sample to test this idea.
In particular, our new $Z-L/L_{\rm Edd}$ relationship may be affected by
a potential selection effect against low accretion rate AGNs at high-$z$,
if such sources are found to be metal-rich.
Our new result is consistent with a scenario in which there is an intimate
relation between AGN fueling, manifested by accretion rate, and
starburst, responsible for the metal-enrichment.
An obvious avenue to proceed is to search for other
evidence of recent star formation in our high-$z$ sample using
IR, mm, and sub-mm observations.

\section{Conclusions}
\label{conclusions}

We present new near-IR spectroscopic measurements of the \hb\
region for a sample of 29 luminous $2<z<3.5$ quasars.
This sample is augmented with measurements of other AGNs (92 AGNs in total) for
which spectroscopic data of the rest-frame UV band as well as the \hb\
region are available. We show that BLR metallicity, indicated by
\nvciv, is primarily correlated with the accretion rate, indicated by
luminosity and FWHM(\hb). We also show that using \ion{C}{4} as a
proxy to \hb\ may be problematic and can lead to spurious BH mass estimates.
The potential implications of our new result to the starburst--AGN
connection and to AGN fueling are briefly discussed.

\acknowledgments

We are grateful to the technical staff at the AAT and TNG
observatories for invaluable help during the observations.
We acknowledge constructive remarks made by an anonymous referee, which helped
to improve this work.
This work is based on observations made with the Italian Telescopio Nazionale
Galileo (TNG) operated on the island of La Palma by the Centro Galileo Galilei
of the INAF (Istituto Nazionale di Astrofisica) at the Spanish Observatorio 
del Roque de los Muchachos of the Instituto de Astrofisica de Canarias.
We would like to thank Angela Cotera for allowing us to use half a night of
her time at AAT, and Dirk Grupe for providing us an electronic version of
the Boroson \& Green (1992) Fe {\sc ii} template.
We gratefully acknowledge constructive remarks from an anonymous referee,
who helped to improve this work considerably.
The 2dF QSO Redshift Survey (2QZ) was compiled by the 2QZ survey team
from observations made with the 2-degree Field on the Anglo-Australian
Telescope.
Funding for the creation and distribution of the SDSS Archive has been
provided by the Alfred P. Sloan Foundation, the Participating
Institutions, the National Aeronautics and Space Administration, the
National Science Foundation, the U.S. Department of Energy, the
Japanese Monbukagakusho, and the Max Planck Society.
The SDSS Web site is http://www.sdss.org/.
The SDSS is managed by the Astrophysical Research Consortium (ARC) for
the Participating Institutions. The Participating Institutions are The
University of Chicago, Fermilab, the Institute for Advanced Study, the
Japan Participation Group, The Johns Hopkins University, Los Alamos
National Laboratory, the Max-Planck-Institute for Astronomy (MPIA),
the Max-Planck-Institute for Astrophysics (MPA), New Mexico State
University, University of Pittsburgh, Princeton University, the United
States Naval Observatory, and the University of Washington.
This research has made use of the NED database which is operated by
the Jet Propulsion Laboratory, California Institute of Technology,
under contract with the National Aeronautics and Space Administration.
This work is supported by the Israel Science Foundation grant 232/03.
RM acknowledges partial support by the Italian Ministry of
Research (MIUR).

\end{document}